 \documentclass[aps, twocolumn, showpacs, amsmath]{revtex4}
\usepackage{dcolumn}% Align table columns on decimal point
\usepackage{bm}% bold math
\usepackage{graphicx}% Include figure files
\usepackage{color}
\usepackage{subfigure}
\usepackage{hyperref}
\usepackage{latexsym}
\usepackage{amsthm}
\usepackage{amssymb}

\DeclareGraphicsExtensions{.jpg,.pdf, .mps, .png, .eps, .ps, .EPS,.gif}

\begin{document}
\def\be{\begin{equation}}
\def\ee{\end{equation}}

\def\bc{\begin{center}}
\def\ec{\end{center}}
\def\bea{\begin{eqnarray}}
\def\eea{\end{eqnarray}}
\newcommand{\avg}[1]{\langle{#1}\rangle}
\newcommand{\Avg}[1]{\left\langle{#1}\right\rangle}

\def\ie{\textit{i.e.}}
\def\etal{\textit{et al.}}
\def\m{\vec{m}}
\def\G{\mathcal{G}}

\newcommand{\davide}[1]{{\bf\color{blue}#1}}
\newcommand{\gin}[1]{{\bf\color{green}#1}}

\title{Synchronization in Network Geometries with Finite Spectral Dimension}

\author{ Ana P. Mill\'an }
\affiliation{Departamento de Electromagnetismo y F\'isica  de la Materia
and Instituto Carlos I de F\'isica Te\'orica y Computacional, Universidad de Granada, 18071 Granada, Spain}
\author{Joaqu\'{\i}n J. Torres}
\affiliation{Departamento de Electromagnetismo y F\'isica  de la Materia
and Instituto Carlos I de F\'isica Te\'orica y Computacional, Universidad de Granada, 18071 Granada, Spain}

\author{ Ginestra Bianconi}
\affiliation{School of Mathematical Sciences, Queen Mary University of London, London, E1 4NS, United Kingdom\\
The Alan Turing Institute, London, United Kingdom}

\begin{abstract}
Recently there is a surge of interest in network geometry and topology. Here we show that  the spectral dimension plays  a fundamental  role in  establishing a clear relation between the topological and geometrical properties of a  network and its dynamics. Specifically we explore the role of the spectral dimension in determining the synchronization properties of  the  Kuramoto model. 
We show that the synchronized phase can only be thermodynamically stable for spectral dimensions above four and that phase entrainment of the oscillators can only be found for spectral dimensions greater than two. We numerically test our analytical predictions on the recently introduced model of network geometry called Complex Network Manifolds which  displays a tunable spectral dimension. 
\end{abstract}


\maketitle

\section{Introduction}

Recently there has been growing interest in characterizing networked structures using geometrical and topological tools \cite{Perspective,Bassett,Lambiotte}. On one side an increasing number of works aim at unveiling the hidden geometry of networks using  statistical mechanics \cite{Emergent,NGF,CQNM,Hyperbolic,Polytopes,Doro_manifold,Boguna1,Boguna2}, discrete geometry \cite{Ollivier_Ricci} and machine learning  \cite{Canistracci,Bruna}, on the other side topological data analysis is tailored to capture the structure of a large variety of network data \cite{Vaccarino,Blue_Brain,Nanoparticles,Lambiotte2,Evans,Aste1,Aste2}.

Simplicial and cell complexes are generalized network structures not only formed by nodes and links but also by triangles, tetrahedra, hypercubes, orthoplexes, etc.  
Having geometrical building blocks, simplicial and cell complexes are ideal discrete structures to investigate and model network geometry and topology \cite{Perspective,Bassett,Lambiotte}.
Modelling network geometry with simplicial and cell complexes has been for long the practice in quantum gravity approaches including Causal Dynamical Triagulations, Regge calculus or Tensor networks, to name a few \cite{Loll,Oriti,Tensor}. 
Moreover, simplicial and cell complexes have recently become very popular to model complex systems ranging from brain networks to social networks  \cite{Perspective,Bassett,Lambiotte,Ana,Petri,Latora}, in part supported by the fact that their geometrical properties are often retained if one considers their network skeleton, i.e. the network formed exclusively by their nodes and links.

Network geometries are typically characterized by having a finite spectral dimension $d_S$ \cite{Toulouse,Burioni,Burioni_universal,Durhuus1,Durhuus2} that characterizes the return time distribution of the random walk. 
For instance, Euclidean lattices in dimension $d$ have spectral dimension $d=d_S$. Therefore in this case the spectral dimension is also equal to the Hausdorff dimension of the lattice, $d_S=d_H$. 
However, in general, networks can have non-integer spectral dimension $d_S$ not equal to their Hausdorff dimension. 
The fundamental role of the spectral dimension in characterizing the geometry of discrete network structures has been widely recognized in quantum gravity where the spectral dimension has been extensively used to compare different approaches \cite{Loll_spectral_dimension,dimensional_reduction,Durhuus1,Durhuus2}.

Interestingly, it has recently been shown that the skeleton of simplicial and cell complexes generated by the model called Complex Network Manifolds \cite{NGF,CQNM,Hyperbolic,Polytopes} displays finite spectral dimension, heterogeneous degree distribution, small-world property (Hausdorff dimension $d_H=\infty$) and rich community structure on top of an emergent hyperbolic geometry.  This suggests that a finite spectral dimension is not only a very strong indication of a rich underlying geometry of network structures, but is also totally compatible with the main universal properties of complex networks.
Therefore, complex networks with a strong geometric component such as brain networks \cite{Sporns,Bullmore,Blue_Brain} and power-grids \cite{Odor}   are likely to display a finite spectral dimension together with characteristic properties of complexity.

Predicting the properties of synchronization dynamics on network geometries is a fundamental statistical mechanics problem that can be crucial to understand  the relation between structural and functional brain networks and to predict the stability of power-grids.
Even though the interplay between complex network structure and synchronization dynamics has been extensively studied \cite{Kurths,Boccaletti,Arenas,Barahona,Villegas,Cota,Lambiotte_XY,kuramoto1975self,strogatz2000kuramoto,acebron2005kuramoto}, so far most works have considered  complex networks  where the smallest non-zero eigenvalue of the Laplacian (the so called Fidler eigenvalue) is well separated from zero, i.e. the network displays a spectral gap and does not display a spectral dimension. 

Only very recently a few works have pointed out that network geometry can have a profound effect on sychronization dynamics \cite{Ana,Blue_Brain,Torre}. 
In particular, it has been found that neuronal cultures have synchronization properties strongly affected by their dimensionality, so that $2d$ neuronal cultures display weaker synchronization properties than neuronal cultures grown in $3d$ scaffolds \cite{Torre}. 
Additionally, large-scale numerical models of the brain generated in the framework of the Blue Brain project \cite{Blue_Brain} reveal that neurons in the brain can be thought of as forming a simplicial complex where neurons belonging to higher dimensional simplices are more correlated. 
Recently, these results have been interpreted in the framework of a numerical stylized model of the Kuramoto model on Complex Network Manifolds displaying strong spatio-temporal fluctuations and strong effects of the dimensionality of the simplicial complex \cite{Ana}. 
 
Here we shed light in these numerical results by investigating the sychronization properties of the Kuramoto model \cite{kuramoto1975self} on networks with a finite spectral dimension. We first derive analytically general results on the predicted stability of the synchronized phase in the linear approximation of the Kuramoto model. Subsequently we compare these predictions with numerical results of the Kuramoto model on Complex Network Manifolds.

For Euclidean lattices of dimension $d$, it is known that the sychronized phase of the Kuramoto model is thermodynamically stable only for $d>4$ \cite{Choi,Chate}. Here we extend this result by showing that in complex networks with finite spectral dimension, the Kuramoto model can yield a synchronized state in the infinite network limit only for spectral dimensions $d_S>4$. 
For spectral dimensions $d_S\in (2,4]$ instead, only an entrained synchronization phase can be  observed in the large network limit. 
Our results are then tested on Complex Network Manifolds formed by regular polytopes of dimension $d$. We validate our results and we show evidence that in these network structures it is possible to observe entrained phase synchronization also for dimensions $d>4$ provided that the spectral dimension $d_S\leq 4$.
Interestingly, Complex Network Manifolds are hyperbolic network geometries \cite{Hyperbolic} which are very different from regular Euclidean lattices. A notable difference with Euclidean lattices is that  despite the fact that they have a finite spectral dimension, their eigenvectors are not delocalized over the network like the Fourier basis on an Euclidean lattice. Rather they can be very localized on a small fraction of nodes, reflecting the symmetries present in the network.
Therefore, here we characterize the spectral properties of Complex Network Manifolds and study the effect of these properties on the entrained phase synchronization, which is known to display strong spatio-temporal fluctuations of the order parameter \cite{Ana}. This phase, also called frustruated synchornization \cite{Villegas,Cota}, has a very rich structure and can be interpreted as an extended critical region to be related to the smeared phase observed in critical phenomena on hyperbolic networks, such as percolation \cite{Ziff,Bianconi_Ziff}.

The paper is organized as follows. In Sec. II we define the properties of the normalized Laplacian and the spectral dimension of a network. 
In Sec. III we introduce the model of Complex Network Manifolds and characterize its spectral properties. 
In Sec. IV we discuss our theoretical  predictions regarding the  synchronization properties of the Kuramoto model on complex networks with finite spectral dimension using the linear approximation. 
In Sec. V we validate the theoretical predictions and fully investigate the properties of synchronization defined over  Complex Network Manifolds. 
In Sec. VI we provide the conclusions. Finally in the Appendices we provide an extensive account of our theoretical derivations.

\section{The spectral dimension}
Diffusion on network structures is typically studied using the properties of suitably defined Laplacian operators.  On an undirected  network of $N$ nodes and adjacency matrix ${\bf a}$ the normalized Laplacian ${\bf L}$ is a $N\times N$ matrix of elements
\bea
L_{ij}=\delta_{ij}-\frac{a_{ij}}{k_i}.
\label{Laplacian}
\eea
The normalized Laplacian operator is typically used to characterize the random walk on a given network, or a diffusion dynamics in which, starting from each node $i$, there is a well defined  probability of diffusion to every neighbour node.
For instance, the random walk can be characterized by studying the equation for the probability $\pi_i(t)$ that a random walker is at node $i$ at time $t$ given by 
\bea
\pi_i(t)=-\sum_jL_{ji}\pi_j(t-1).
\eea
Given the initial condition $\pi_i(0)=\delta_{i,i_0}$, this equation has the solution
\bea
\pi_i(t)=\sum_{\lambda}e^{-\lambda t}u_i^{\lambda}v_{i_0}^{\lambda},
\eea
where ${\bf v}^{\lambda}$ and ${\bf u}^{\lambda}$ are the right and left eigenvectors corresponding to the eigenvalue $\lambda$.
While ${\bf L}$ is asymmetric, an alternative definition of the normalized Laplacian considers the symmetric matrix ${\bf \hat{L}}$ of elements
\bea
\hat{L}_{ij}=\delta_{ij}-\frac{a_{ij}}{\sqrt{k_ik_j}}.
\eea
Interestingly, is it easy to show that the spectrum of ${\bf L}$ and the spectrum of ${\bf \hat{L}}$ are the same. Therefore, although the normalized Laplacian ${\bf L}$ is asymmetric, it has a real spectrum and non-negative eigenvalues. 
Additionally, the normalized Laplacian has the following spectral properties:
\begin{itemize}
\item
 The normalized Laplacian {\bf L} has always one zero eigenvalue $\lambda=0$ with degeneracy equal to the number of components of the network. So if a network is connected the zero eigenvalue has degeneracy one. 
 \item
 In a connected network the right and left eigenvectors corresponding to the zero eigenvalue $\lambda=0$ are given by 
\bea
{\bf v}^{\lambda=0}&=&\frac{1}{\sqrt{\avg{k}N}}(1,1,\ldots 1), \nonumber \\
{\bf u}^{\lambda=0}&=&\sqrt{\avg{k}N}(\mu_1,\mu_2,\ldots, \mu_N),
\label{eigen}
\eea
where 
\bea
\mu_i=\frac{k_i}{\Avg{k}N}
\eea
is the invariant measure of the random walk on the network.
The components of the right ${\bf v}^{\lambda}$ and left ${\bf u}^{\lambda}$ eigenvectors of ${\bf L}$ are related to the components of the eigenvectors ${\bf w}^{\lambda}$ of $\hat{\bf{L}}$ by  
\bea
u_i^{\lambda}&=&\sqrt{{k_i}}w_i^{\lambda}, \nonumber \\
v_i^{\lambda}&=&\frac{1}{\sqrt{k_i}}w_i^{\lambda}.
\eea
Therefore, if follows that the elements $u_i^{\lambda}$ and $v_i^{\lambda}$ are simply related by the expression
\bea
u_{i}^{\lambda}=k_iv_{i}^{\lambda}.
\eea
Moreover, since the eigenvectors ${\bf w}^{\lambda}$ are orthogonal, we have 
\bea
\sum_{i=1}^N u_{i}^{\lambda}v_{i}^{\lambda^{\prime}}=\sum_{i=1}^N w_i^{\lambda}w_i^{\lambda^{\prime}}=\delta(\lambda,\lambda^{\prime}).
\label{norm}
\eea
\item
The effective number of nodes over which the $\lambda$ eigenmode is localized can be measured using the {\it participation ratio} $Y$ defined as \cite{Ana}
\bea
Y&=&\left[\sum_{i=1}^N (u_i^{\lambda}v_i^{\lambda})^2\right]^{-1}\nonumber\\
&=& \left[\sum_{i=1}^N (w_i^{\lambda})^4\right]^{-1}.
\eea
\end{itemize}
In networks with distinct geometrical properties, the density of eigenvalues $\rho(\lambda)$ of the normalized Laplacian follows the scaling relation
\bea
\rho(\lambda)\simeq \lambda^{d_S/2-1}
\label{scaling}
\eea
for $\lambda\ll1$, where $d_S$ is called the {\it spectral dimension} of the network.
In $d$-dimensional Euclidean lattices $d_S=d$. More generally, it can be shown that $d_S$ is related to the Hausdorff dimension $d_H$ of the network by the disinequalities \cite{Durhuus1,Durhuus2}
\bea
d_H\geq d_S\geq 2\frac{d_H}{d_H+1}. \label{eq:dHdS}
\eea
Therefore, for small-world networks, which have infinite Hausdorff dimension $d_H=\infty$, it is only possible to have finite spectral dimension $d_S\geq 2$.

We observe here that, in presence of a finite spectral dimension, the cumulative distribution $\rho_c(\lambda)$ evaluating the density of eigenvalues $\lambda'\leq \lambda$ follows the scaling 
\bea
\rho_c(\lambda)\simeq \lambda^{d_S/2}, 
\label{eq:rho_c}
\eea
for $\lambda\ll 1$. In presence of a finite spectral dimension it is possible to evaluate the scaling with the network size of the  smallest non-zero eigenvalue $\lambda_2$ of a connected network (also called the  the Fidler eigenvalue) by imposing that 
\bea
\rho_c(\lambda_2)=\frac{1}{N},
\eea 
i.e. the eigenvalue $\lambda_2$ is the smallest non zero eigenvalue.
From this relation and the scaling of the cumulative density of eigenvalues we get
\bea
\lambda_{2}\propto N^{-2/d_S}.
\label{Fidler}
\eea
Therefore, the Fidler eigenvalue $\lambda_2\to 0$ as $N\to \infty$ and we say that in the large network limit the spectral gap closes.

\section{Complex Network Manifolds: a model with tunable spectral dimension}
\subsection{Definition and basic structural properties}

Simplicial complexes and cell complexes are natural objects to be considered when investigating network geometry. In fact,  they can be intuitively interpreted as geometrical network structures built from geometrical building blocks. 

A pure $d$-dimensional simplicial complex is formed by $d$-dimensional simplices (fully connected networks of $d+1$ nodes) such as nodes ($d=0$), links $(d=1)$, triangles $(d=2)$ , tetrahedra $(d=3)$ etc., glued along their faces.
Here by a face of a $d$-dimensional simplex, we indicate a $\delta$-dimensional simplex with $\delta<d$ formed by a subset of its nodes.
A simplicial complex has the following two  additional properties:
\begin{itemize}  
\item[(1)]
If a simplex $\alpha$ belongs to the simplicial complex $\mathcal K$  (i.e. $\alpha\in {\mathcal K}$), then also all its faces $\alpha'\subset \alpha$ belong to the simplicial complex $\mathcal K$ (i.e. $\alpha'\in {\mathcal K}$).
\item[(2)]
If two simplices $\alpha$ and $\alpha'$ belong to the simplicial complex (i.e. $\alpha,\alpha'\in {\mathcal K}$), either their intersection is null, i.e. $\alpha \cap \alpha'= \emptyset$ or their intersection belongs to the simplicial complex, (i.e. $\alpha \cap \alpha'\in {\mathcal K}$).
\end{itemize} 

Here we consider a recently proposed model, Complex Network Manifolds (CNM) \cite{CQNM,NGF,Hyperbolic}, that generates discrete $d$-dimensional manifolds by a non-equilibrium growing simplicial complex dynamics.
CNM are discrete manifolds  generated by gluing subsequently $d$-dimensional simplices along their $(d-1)$-faces.
Every $(d-1)$-face $\alpha$ of the CNM is characterized by an incidence number $n_{\alpha}$ indicating the number of $d$-dimensional simplices incident to it minus one.
Initially (at time $t=1$), the CNM is formed by a single $d$-dimensional simplex.
At any subsequent step (at time $t>1$), a new $d$-dimensional simplex is glued to a $(d-1)$- face $\alpha$ with probability
\bea \Pi_{\alpha} = \frac{ 1-n_\alpha}{\sum_{\alpha'}(1-n_{\alpha'})}.\label{prob}\eea
In Ref. \cite{Hyperbolic} the exact degree distribution of CNM has been analytically  derived.  Mainly, the degree distribution $\tilde{P}(k)$ is exponential  for dimension $d=2$ and power-law (i.e. $\tilde{P}(k)\simeq Ck^{-\gamma}$) for dimension $d>2$, with power-law exponent $\gamma$ given by 
\bea 
 \gamma=2+\frac{1}{d-2}.
 \eea
 
CNM can be generalized to cell complexes that are not just formed by simplices but instead they are formed by the subsequent gluing of regular polytopes along their faces \cite{Polytopes}.
Since in dimension $d>4$ there are only three types of convex regular polytopes, the simplices, the hypercubes and the orthoplexes, here we focus on CNM formed by subsequently gluing these building blocks along their faces. 
Therefore, we consider CNM built using repeatedly the same building block given by a $d$-dimensional simplex, a $d$-dimensional hypercube or a $d$-dimensional orthoplex.
To each face of the polytopes we assign an incidence number $n_{\alpha}$ given by the number of $d$-dimensional polytopes incident to it minus one. Finally the cell complex is built by starting from a single polytope and at each subsequent time adding a new polytope of the same type to a $(d-1)$-face with probability given by Eq. $(\ref{prob})$.
 
The resulting CNM \cite{Polytopes} have exponential degree distribution for $d=2$ and power-law degree distribution for $d>2$, with power-law exponent $\gamma $ given by  
\bea
   \gamma = 1 + \frac{F-2}{f-2},
   \label{gamma}
\eea
where $F$ is the number of faces of the regular polytopes that form the building block of the cell complex, and $f$ is the number of $(d-1)$-faces incident to a node on the same regular polytope.
By using the fact that $F$ and $f$ are given for the different regular polytopes by 
\bea
\begin{array}{lll}
F=d+1, &f=d,& \mbox{simplices},\\
F=2d,& f=d, & \mbox{hypercubes},\\
F=2^d, &f=2^{d-1},& \mbox{orthoplexes},
\end{array}
\eea
we derive that the power-law exponent $\gamma$ of the degree distribution is given by 
\bea
\begin{array}{ll}
\gamma=2+\frac{1}{d-2},&\mbox{simplices},\\
\gamma=3+\frac{2}{d-2},&\mbox{hypercubes},\\
\gamma=3+\frac{1}{2^{(d-2)}-1},& \mbox{orthoplexes}.
\end{array}
\eea
Interestingly, we notice that only CNM built using simplices have a scale-free degree distribution with $\gamma\in(2,3]$ in dimension $d>2$.

We observe that the  network structure of simplicial complexes CNM \cite{CQNM} reduces to Apollonian Random Graphs \cite{apollonian1,apollonian2} and the cell complexes CNM in $d=3$ are strictly related to the model proposed in Ref. \cite{Aste}.

It was recently revealed that CNM and their generalization called Network Geometry with Flavor \cite{NGF}, which allows us to establish the connection with preferential attachment models, have an emergent hyperbolic geometry. 
Here we focus exclusively on the skeleton of CNM, i.e. the network formed exclusively by its nodes and links.
The geometrical nature of the skeleton of CNM is strongly reflected in the spectral properties of the network, characterized by a finite spectral dimension, as we will discuss in the following section.

\subsection{The spectral properties of Complex Network Manifolds}
CNM follow simple combinatorial rules that do not take into account any embedding space. However, these structures display an emergent hyperbolic geometry characterized by an infinite Hausdorff dimension $d_H=\infty$ (the networks are small-world) \cite{Ana} together with a finite spectral dimension $d_S\geq 2$. 

\begin{figure}
\begin{center}
\includegraphics[width=1.0\columnwidth]{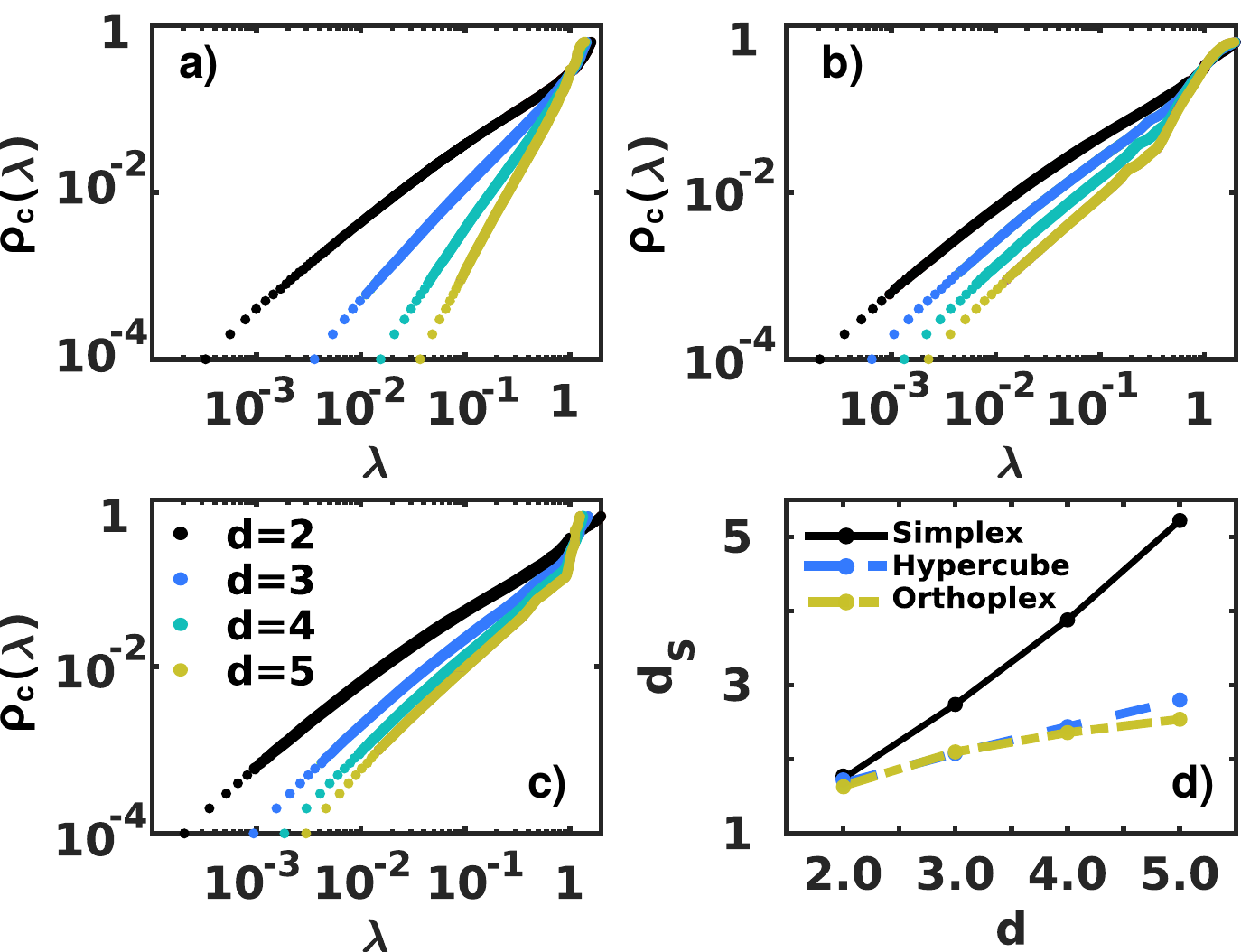} 
\caption{  The cumulative distribution of eigenvalues
$\rho_c(\lambda)$ for CNM of dimension $d = 2,\ 3,\ 4$ and $5$, is shown in panels $(a)$, $(b)$ and $(c)$  for the simplex, hypercube and orthoplex CNM respectively. Panel $(d)$ represents the fitted spectral dimension of the CNM as a function of the dimension $d$ of its building blocks. Results are for $ N = 6400$ and the cumulative distribution of eigenevalue $\rho_c(\lambda)$ is averaged over $100$ realizations of the network. \label{fig:F1} }
\end{center}
\end{figure}

In this section we investigate numerically the spectral properties of CNM. 
Figure $\ref{fig:F1}$ shows the cumulative distribution of eigenvalues $\rho_c(\lambda)$ as obtained for the simplices (panel $a$), hypercubes (panel $b$) and orthoplexes (panel $c$), and for dimensions $d=2,\ 3,\ 4$ and $5$, as indicated by the different colours in the legend. 
A finite size study of this spectrum reveals that $\lambda_2$ approaches zero in the large network limit, as predicted in presence of a finite spectral dimension $d_S$. 
Moreover, $\rho_c(\lambda)$ obeys Eq.($\ref{eq:rho_c}$) for $\lambda\ll 1$, which allows us to obtain the spectral dimension $d_S$ as a function of $d$ (see Figure $\ref{fig:F1}d$) by performing a power-law fit to $\rho_c(\lambda)$ for $\lambda\ll1$.
We notice that the spectral dimension $d_S$ increases with the dimension of the regular polytope $d$ for simplices, hypercubes and orthoplex as well. However, the growth of $d_S$ with $d$ saturates for hypercubes and orthoplexes, while it does not appear to saturate for simplices. 
Therefore, we conclude that the spectral dimension $d_S$ does not only depend on the dimension $d$ of the polytopes forming the building blocks of the cell complex, but also on the specific  nature and symmetry of these polytopes. 

\begin{figure*}
\begin{center}
\includegraphics[width=1.6\columnwidth]{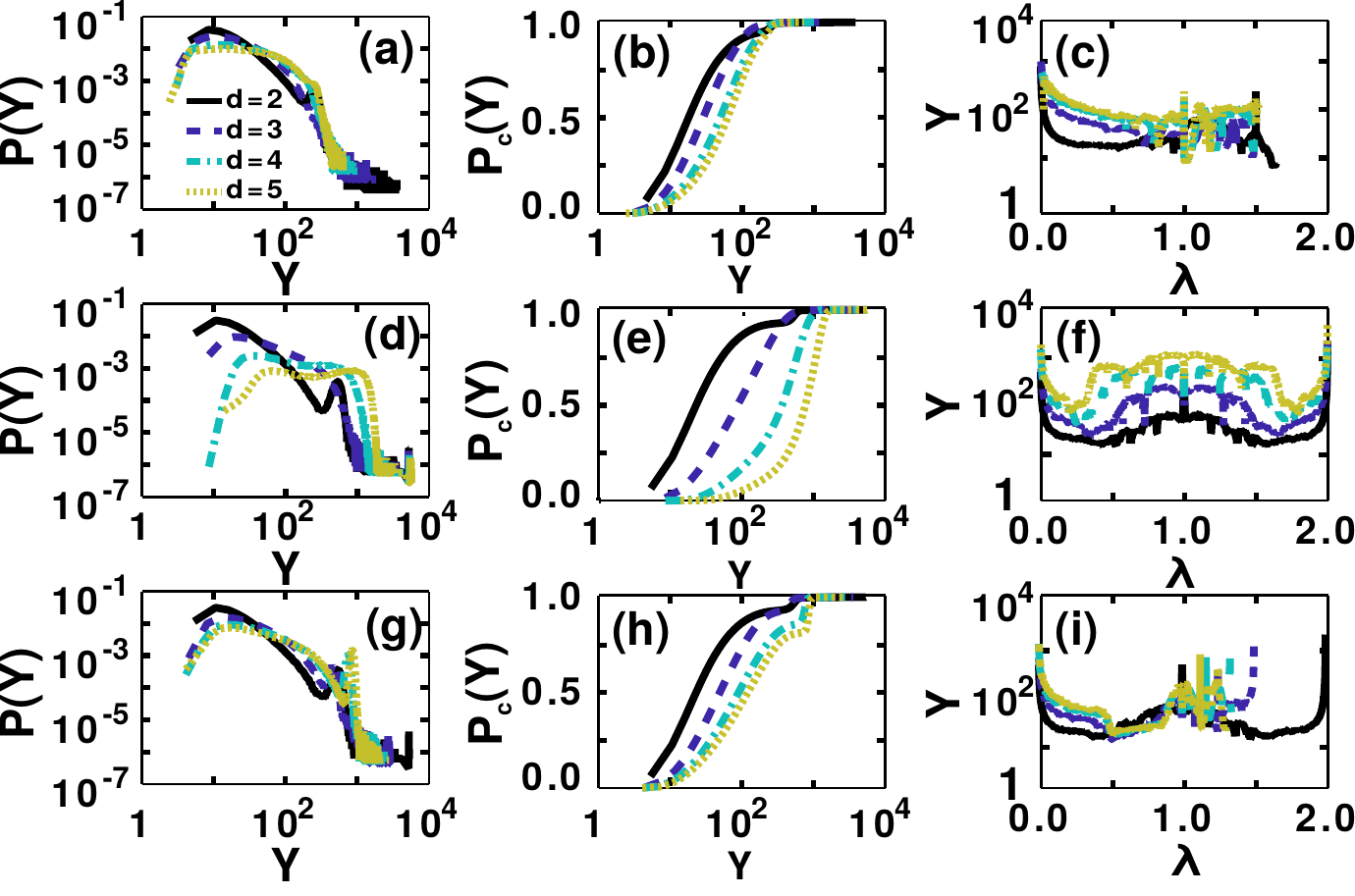} 
\caption{ The probability distribution $P(Y)$, the cumulative distribution $P_c(Y)$ of the participation ratio $Y$, and the average value of the participation ratio $Y$ as a function of the corresponding eigenvalue $\lambda$ are shown for CNM formed by  simplices (panels $(a,b,c)$), hypercubes (panels $(d,e,f)$) and orthoplexes (panels $(g,h,i)$) networks with dimension $d=2,3,4,5$ of the polytopes. \label{fig:F2} }
\end{center}
\end{figure*}

Moreover, we observe that although CNM appear to have a finite spectral dimension as Euclidean lattices, the eigenvectors of CNM are very different from the Fourier eigenvectors of a Euclidean lattice, as evidenced by the behavior of its participation ratio $Y$ (see Figure $\ref{fig:F2}$). 
In fact, for Euclidean lattices one would have $Y=N$ for all eigenmodes, while for CNM there is a large fraction of eigenmodes with partion ratio $Y\ll N$.  
The eigenvectors have indeed a very heterogeneous distribution $P(Y)$ of the participation ratio $Y$, including many eigenvectors localized on a small number of nodes compared to the total number of nodes of the network (see panels $(a)$, $(d)$, and $(g)$ of Figure $\ref{fig:F2}$). 
This phenomenon can be also appreciated by observing that the cumulative distribution $P_c(Y)$ of eigenmodes with partition ratio less than $Y$ can be significantly high also for values of $Y$ much smaller than the number of nodes $N$ of the network, i.e. $Y\ll N$ (see panels $(b)$, $(e)$, $(h)$ of Figure $\ref{fig:F2}$).
Finally, the dependence of the participation ratio $Y$ on $\lambda$ can be highly non-trivial (panels $(c)$, $(f)$, and $(i)$ of Figure $\ref{fig:F2}$) and it is likely to be affected by the symmetries of the CNM \cite{Sanchez}.

\section{ Kuramoto dynamics on networks with finite spectral dimension}

\subsection{The Kuramoto model}

Synchronization dynamics on complex networks has been widely studied in the literature and it is known to be very significantly affected by the spectral properties of the network. However, the scientific interest so far has focused on networks which do not have a spectral dimension and display instead what is called a spectral gap, i.e. the smallest non-zero eigenvalue of the normalized Laplacian $\lambda_2$ does not approaches zero in the infinite network limit.

However, in network geometries it is important to consider network structures in which the spectral gap closes, $\lambda_2\to 0$ as $N\to \infty$, and the density of eigenvalues follows the scaling in Eq. ($\ref{scaling}$), i.e. the network has a finite spectral dimension. To investigate the role of the spectral dimension in the synchronization dynamics, we consider the Kuramoto model.

The Kuramoto dynamics describes a system of $N$ coupled oscillators $i=1,2,\ldots, N$ with phases $\theta_i(t)$ obeying the following dynamical equation,
\begin{equation}
\dot{\theta}_i(t)=\omega_i + \sigma\sum_{j=1}^N  \frac{a_{ij}}{k_i} \sin (\theta_j - \theta_i),
\label{Kuramoto}
\end{equation}
where $k_i$ is the degree of node $i$, $a_{ij}$ the adjacency matrix of the network, and $\sigma$ the control parameter tuning the strength of the coupling between nodes. 
Each internal frequency $\omega_i$ is independently drawn from a normal distribution with mean $0$ and variance $1$, i.e. ${\mathcal N}(0,1).$
We note that sometimes the Kuramoto model is defined by omitting $k_i$ in Eq. ($\ref{Kuramoto}$), however our choice here is dictated by the desire to screen out the effect of having heterogeneous degree distributions. 
Therefore, the considered dynamics is designed to be independent of the degree distribution so that the effect of having networks with different spectral dimension can be revealed.

\subsection{Theoretical predictions}
In order to study the stability of the synchronized phase, we have linearized the Kuramoto dynamics in Eq. $(\ref{Kuramoto})$ assuming that $|\theta_i-\theta_j|\ll1$ for every pair of neighbour nodes. 
In this way we get the linear system of equations 
\bea
\dot{\theta}_i(t)=\omega_i - \sigma\sum_{j=1}^N  {L}_{ij}\theta_j,
\label{linearized}
\eea
for $i=1,2,\ldots, N$, where $\bf {L}$ is defined in Eq. ($\ref{Laplacian}$).
In order to evaluate the stability of the  synchronized state, we use an approach already established for finite lattices \cite{Choi,Chate}. Specifically  we calculate the average fluctuation of the phases over the entire network by evaluating $W^2$ given by 
\bea
W^2=\frac{1}{N}\Avg{\sum_{i=1}^N[\theta_i(t)-\overline{\theta}]^2},
\label{W2a}
\eea
 where in Eq. (\ref{W2a})  $\overline{\theta}$ is given by
\bea
\overline{\theta}= \frac{1}{N}\sum_{i=1}^N{\theta_i(t)},
\eea
in the linear approximation.
In presence of a thermodynamically stable synchronized phase, the average fluctuations of the phases $W^2$ should remain bounded. Therefore, if $W^2$ diverges with the network size $N$, the  synchronized phase is unstable.
By considering networks having a finite spectral dimension $d_S$ we obtain (see Appendix $\ref{a1}$) that in the large network limit ($N\to \infty$) $W^2$ diverges as long as $d_S\leq 4$. Specifically we can show that $W^2$ obeys the scaling
\bea
     W^2 \sim  \left\{\begin{array}{ll}N ^{4/{d}_S-1} & \mbox{if }  {d}_S < 4, \\
		\ln(N) & \mbox{if } {d}_S = 4 ,\\
		\mbox{const} & \mbox{if } {d}_S > 4.
	\end{array}\right.
\eea
It follows from this derivation that the  synchronized state cannot be thermodynamically stable in networks with spectral dimension $d_S\leq 4$.

The linear approximation is valid only if the  coupling term of each oscillator  with the  phases of the linked oscillators is small. Therefore in order for the linear approximation to hold we must require  that the vector ${\bf L}\bm{\theta}$ has small elements.
 A global parameter that can establish the sufficient condition for the failure of  the linear approximation is the {\it correlation}  $C$ defined as 
\bea
    C= \frac{1}{N}\Avg{\bm{\theta}^{T} {\bf L}\bm{\theta}}.
\eea
In fact, if the correlation $C$ diverges the linear approximation cannot be valid.
In a network with finite spectral dimension $d_S$ we have obtained (see detailed derivation in Appendix $\ref{a2}$) that $C$ obeys the following scaling with $N$,
\bea
C\sim  \left\{
	\begin{array}{ll}
		 N ^{2/{d}_S-1} & \mbox{if }  {d}_S < 2, \\
		\ln(N) & \mbox{if } {d}_S = 2 ,\\
		\mbox{const} & \mbox{if } {d}_S > 2.
	\end{array}\right.
\eea
Therefore, for spectral dimension $d_S\leq 2$ the correlations among the phases of nearest neighbour nodes diverge and the linear approximation fails.

So far we have shown that for spectral dimension $d_S< 2$ the linear approximation fails, while for spectral dimensions $d_S\in (2,4]$ the linear approximation can be valid but the  synchronized phase is not  thermodynamically stable. In order to uncover the phenomenology for spectral dimensions $d_S \in (2,4]$, we follow the approach used by \cite{Choi,Chate} for regular lattices.
We start by characterizing the fluctuations observed in phase velocities across the nodes of the network
\bea
V^2=\frac{1}{N}\sum_{i=1}^N\Avg{\left[{\psi}_i-\bar{\psi}\right]^2},
\eea
where $\psi_i$ indicates the phase velocity of node $i$, 
\bea
\psi_i=\dot{\theta_i},
\eea
and $\bar{\psi}$ the average of the phase velocities over the network
\bea
\bar{\psi}=\frac{1}{N}\sum_{i=0}^N\psi_i. 
\eea
In Appendix \ref{a3} we show that, as long as the linear approximation is valid, i.e. $d_S>2$,
the fluctuations observed in phase velocities vanish in the large network limit, i.e.
\bea
V^2\to 0 & \mbox{as} & N\to \infty.
\eea
This analysis therefore reveals that for spectral dimensions $d_S \in (2,4]$ phase entrainment takes place as long as the linear approximation is valid.

\section{Kuramoto model on Complex Network Manifolds}

\begin{figure*}[htb!]
\begin{center}
\includegraphics[width=1.8\columnwidth]{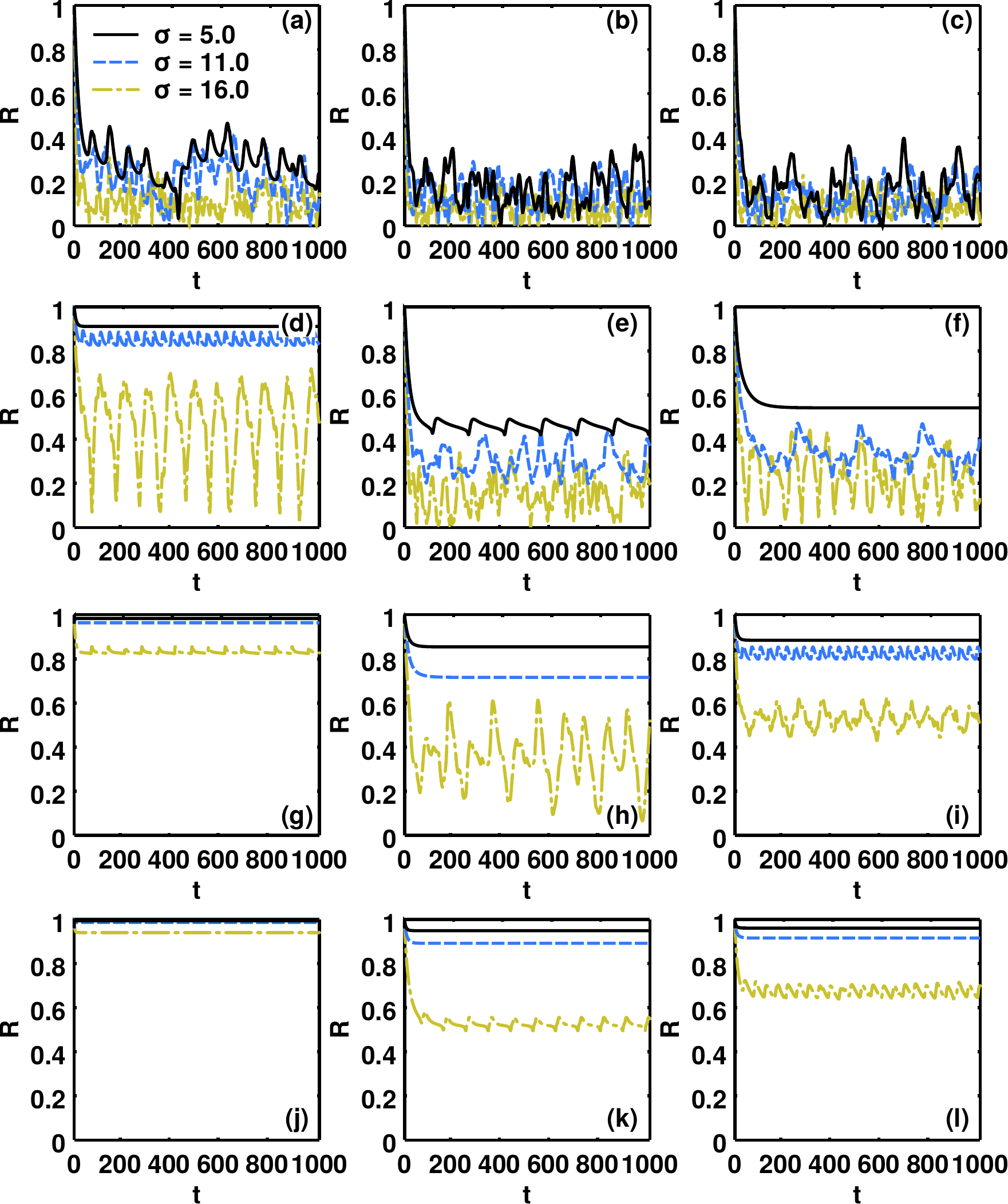} 
\caption{Time series of the global order parameter calulcated for different values of $\sigma=5$, $11$ and $16$, as indicated in the legend, and for CNM formed by simplices (panels $(a)$, $(d)$, $(g)$, $(j)$), hypercubes (panels $(b)$, $(e)$, $(h)$, $(k)$) and orthoplex (panels $(c)$, $(f)$, $(i)$, $(l)$) of dimensions $d=2$ (panels $(a)$, $(b)$, $(c)$), $d=3$ (panels $(d)$, $(e)$, $(f)$), $d=4$ (panels $(g)$, $(h)$, $(i)$) and $d=5$ (panels $(j)$, $(k)$, $(l)$). \label{fig:series} }
\end{center}
\end{figure*}

In this section we present numerical results of the  Kuramoto dynamics defined over CNM. As CNM have tunable spectral dimension this analysis will provide a solid benchmark where we can test our theoretical predictions.
The macroscopic state of synchronization of the system at each time $t$ is characterized by the Kuramoto order parameter, defined as
\begin{equation}
Z(t) = R(t) e^{i\phi(t)} = \frac{1}{N} \sum_{j=1} ^N e^{i\theta_j(t)},
\end{equation}
where $R(t) \in \left[ 0,1 \right]$ is a real variable that quantifies the level of global synchronization, and $\phi(t)$ gives the average global phase of collective oscillations \cite{Kurths,kuramoto1975self}. 
Therefore, $R(t) \approx 0$ corresponds to the noisy or non-coherent state, whereas $R(t) \approx 1$ corresponds to  the coherent or synchronized state. 

We simulated the Kuramoto dynamics by integrating the system of Eqs. ($\ref{Kuramoto}$) in  \textit{MATLAB} using the ode45 function, which uses a non-stiff $4$-th order integration algorithm with adaptive time steps.  Simulations are run for a total time $T$, and for different realizations of the CNM, formed by $d$-dimensional simplices, hypercubes and orthoplexes.
\begin{figure*}
\begin{center}
\includegraphics[width=2.0\columnwidth]{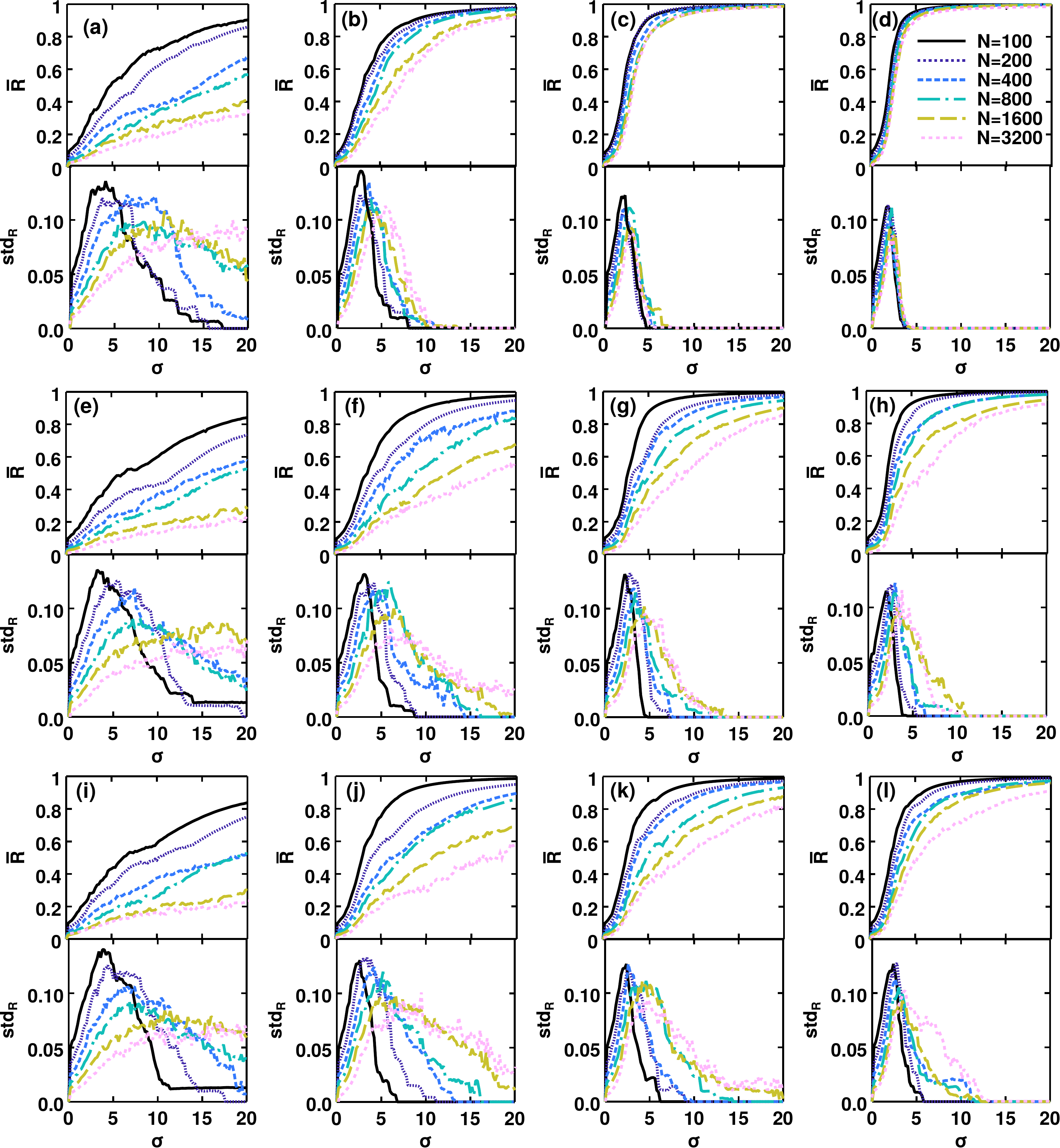} 
\caption{The average order parameter $\bar{R}$ and the standard deviation of the order parameter $std_R$) are plotted versus the coupling constant $\sigma$ for CNM formed by simplices (panels $(a)$, $(b)$, $(c)$, $(d)$), hypercubes (panels $(e)$, $(f)$, $(g)$, $(h)$) and orthoplexes (panels $(i)$, $(j)$, $(k)$, $(l)$) and for dimension $d=2$ (panels $(a)$, $(e)$, $(i)$), $d=3$ (panels $(b)$, $(f)$, $(j)$), $d=4$ (panels $(c)$, $(g)$, $(k)$) and $d=5$ (panels $(d)$, $(h)$, $(l)$). 
Results are shown for different network sizes $N=100,\ 200,\ 400,\ 800,\ 1600$ and $3200$ as indicated in the legend. Results are for $T = 1000$ and have been averaged after equilibration  for $20$ realizations of the networks and internal frequencies. \label{fig:F3} }
\end{center}
\end{figure*} 

Our numerical analysis reveals that CNM can display a frustrated synchronization phase with fully  entrained  phases in which the global order parameter $R(t)$ has large temporal fluctuations. 
The typical range of values of the coupling constants where we observe this phase depends both on the spectral dimension $d_S$ and the network size $N$. In  Figure $\ref{fig:series}$ we show single instances of the time series $R(t)$ of the global order parameter defined on CNM of size $N=3200$ for representative values of the coupling $\sigma$ and for different polytopes and dimensions. Characteristic states of frustrated synchronization  can be observed for $\sigma = 5.0$  and $d=3,4$ for CNM formed by simplex, hypercubes and orthoplex (panels $(d)$, $(e)$, $(f)$, $(g)$, $(h)$ and $(i)$ of Figure ${fig:series}$); and also for $d=5$ for CNM formed by hypercubes and orthoplex (panels $(k)$ and $(l)$ for Figure $\ref{fig:series}$).

In general, for CNM formed by a finite number of nodes $N$, as the coupling constant $\sigma$ increases we can generally distinguish between three phases. 
For very small values of the  coupling constant $\sigma$, the order parameter $R(t)\approx 0$, i.e. the oscillators are not coherent (as shown for example in panels $(a)$ and $(b)$ of Figure $\ref{fig:series}$). 
For large values of the coupling constant $\sigma$ we observe a synchronized phase and a stationary time-series of $R(t)$ with large values of $R(t)$ (see for instance curves obtained for $\sigma = 11.0$,  and $\sigma = 16.0$, in panels $(a)$ and $(b)$ of Figure $\ref{fig:series}$). 
In the intermediate range of values of the coupling constant $\sigma$, we observe the frustrated synchronization regime of phase entrainment  where the order parameter $R(t)$ is not stationary (see for instance curves obtained for $\sigma = 5.0$  in panels $(d)$, $(e)$ and $(f)$ of Figure $\ref{fig:series}$).

\begin{figure*}[htb!]
\begin{center}
\includegraphics[width=2.0\columnwidth]{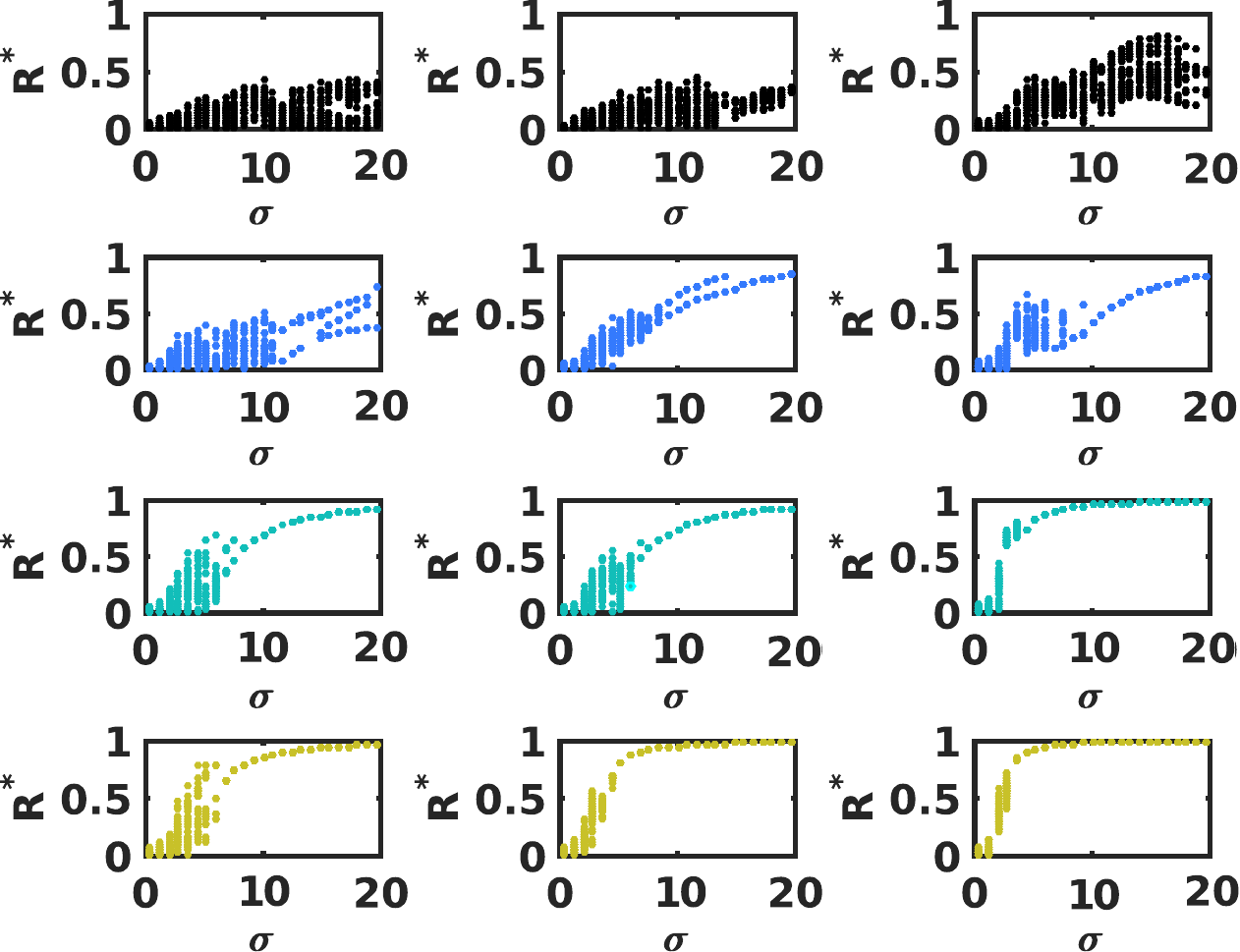} 
\caption{
%Poincar\'e section of
Orbit diagrams of the system dynamics for CNM formed by simplices (left panels), hypercubes (center panels) and orthoplexes (right panel) for $d=2,\ 3,\ 4$ and $5$ from top to bottom. 
%The Poincar\'e section is 
The orbit diagrams are represented by the extremes (maxima and minima) $R^*$ taken by $R(t)$ for $t>0.8T$. Results are for $N=3200$, $T=1000$ and a given realization of the networks structure. } \label{fig:F4} 
\end{center}
\end{figure*}

In order to investigate the thermodynamical stability of these phases in the large network limit as a function of the spectral dimension $d_S$, we have studied the finite size effects of the Kuramoto synchronization for CNM formed by simplices, hypercubes and orthoplexes for dimensions $d=2,3,4,5$.
The spectral dimension of these CNM is shown in Figure $\ref{fig:F1}$.
For $d=2$ CNM have spectral dimension $d_S\approx 2$, whereas for $2<d\leq 5$ CNM formed by simplices have spectral dimension $d_S$ that in first approximation can be assumed to be $d_S\approx d$ and CNM formed by hypercubes and orthoplexes have spectral dimension $d_S\in (2,3)$.
Consequently, our theoretical expectation is that for $d=2$ we cannot observe entrained phases, and that for $2<d\leq 4$ we can observe entrained phases and the synchronized phase cannot be thermodynamically stable. Moreover  for $d=5$ our predictions are that CNM formed by simplices can display a thermodynamically stable  synchronized phase while CNM formed by hypercubes and orthoplexes cannot display a thermodynamically stable  synchronized phase.

In order to test these predictions we have numerically studied as functions of the coupling $\sigma$ the   mean value $\bar{R}$ and the standard deviation $std_R$ of the  order parameter $R(t)$, averaged after the transient evolution over different realizations of CNM. 
In Figure $\ref{fig:F3}$ we display $\bar{R}$ and $std_R$ for CNM formed by simplices, hypercubes and orthoplex of dimension $d=2,3,4,5$ and different network sizes $N$.
The de-coherent or unsynchronized phase corresponds to the regime where $\bar{R}$ is low. The synchronized phase corresponds to the regime where  $\bar{R}$ is high and the fluctuations $std_R$ are low. Finally, the frustrated synchronization phase corresponds to values of the coupling where both $\bar{R}$ and $std_R$ have significantly high values.
As the network size $N$ increases we observe different scenarios depending on the value of the spectral dimension $d_S$.
For spectral dimension $d_S\approx 2$, in the large network limit the system remains in the de-coherent state. This occurs for all considered CNM of dimension $d=2$.
For spectral dimension $d_S\in (2,4]$, we observe that the  synchronized phase is not thermodynamically stable as the values of  coupling constant where the onset of this phase is  observed increase with the network size and do not converge to a finite value. It occurs for CNM formed by simplices of dimension $d=3,4$ and for CNM formed by hypercubes and orthoplexes of dimension $d=3,4,5$. 
Finally, for spectral dimension $d_S>4$ we observe that the synchronized phase is thermodynamically stable as the onset of this phase occurs at a finite value of $\sigma$ in the large network limit.

In summary, our numerical study of  the synchronization properties of CNM indicates that the phase diagram of the model depends critically on the spectral dimension $d_S$ as predicted by our theoretical investigation.

The properties of the frustrated synchronization phase observed in CNM are here furthermore investigated by means of the 
%Poincar\'e section 
orbit diagrams \cite{Poincare} (see  Figure $\ref{fig:F4}$).  These are measured as the extrema $R^*$ (maximum and minimum) of the time series $R(t)$ for each coupling $\sigma$. %, and are an extension of the Poincar\'e section to 1D systems.
Therefore, a fixed stationary state is represented  by one point corresponding to the mean value, as it appears  in the synchronized state observed for high values of $\sigma$ provided that $d>2$. This situation corresponds to one of full synchronization if $R^*=1$ or to partial synchronization if $R^*<1$, in which some nodes remain unsynchronized. 
For  spectral dimensions $d_S\in (2,4]$, on the other hand, we observe that, as the value of the coupling constant $\sigma$ is lowered and we enter in the frustrated synchronization phase,  oscillatory states appear with a given number of extrema that depends on the network and frequency realization. 
These typically  correspond to intereference among different locally synchronized regions, whose sizes scale as $N$ \cite{Ana}, which gives rise to a chaotic behavior as the coupling constant $\sigma$ is decreased. Finally, in the case $d_S\approx d=2$ the synchronized state is never reached.

\section{Conclusions}

This work investigates  the role of the spectral dimension $d_S$ on the synchronization properties of the Kuramoto model. Using  a linear approximation we have shown that the  synchronized phase cannot be thermodynamically stable for spectral dimension $d_S\leq 4$. 
Therefore a necessary condition to observe a  synchronized regime in the thermodynamic limit is that $d_S>4$.  
We have also shown that the considered linear approximations cannot be valid for $d_S\leq2$, since the correlations $C$ diverge. 
Finally, we have shown that, for spectral dimension $d_S \in \left( 2,4 \right]$, phase entrainment takes place in the large network limit as long as the linear approximation is valid, i.e. the fluctuations in phase velocities, $V^2$, vanish asymptotically in time, so that the phases of the nodes are totally entrained.

In order to consider a concrete  example where to test these theoretical derivations, we have characterized the synchronization dynamics of the normalized Kuramoto model taking place on Complex Network Manifolds which have a tunable spectral dimension. 
These networks define discrete manifolds with the small-world property (infinite Hausdorf dimension) and highly modular structure, and provide an ideal theoretical setting to explore the interplay between network geometry and synchronization dynamics \cite{Ana}. 

CNM have significant spectral properties and display a finite spectral dimension. In particular, we have found that CNM based on simplicial complexes have a spectral dimension $d_S$ increasing almost linearly with the dimension $d$ of the simplices, whereas CNM formed by $d$-dimensional hypercubes and orthoplexes have a spectral dimension $d_S$ that saturates with $d$. 
Having a tunable spectral dimension, CNM can be compared to Euclidean lattices that have a spectral dimension $d_S$ equal to their Hausdorff dimension, i.e. $d_S=d_H$. 
However, CNM have a hyperbolic structure with $d_H=\infty$ and we always observe $d_S<d_H$.
Moreover, a closer look at the localization properties of the eigenvectors CNM reveals more significant differences with respect to Euclidean lattices. 
In fact, contrarily to the Fourier eigenvector of Euclidean lattices, a large fraction of eigenmodes of CNM are highly localized on few nodes of the network, reflecting the symmetries of the building block structure.

We have studied numerically the Kuramoto dynamics on CNM testing our theoretical predictions on the nature of the  synchronization dynamics as a function of the spectral dimension $d_S$.  We show that a  frustrated synchronization regime with  entrained  phases emerges for spectral dimensions $d_S\in (2,4]$ and that, for this range of values of the spectral dimension, finite CNM with high coupling constant $\sigma$ reach also a  synchronized phase but this phase is not thermodynamically stable. Moreover, we show that for spectral dimension $d_S=5$ the  synchronized phase is thermodynamically stable.

In conclusion our work  reveals that non-trivial synchronization states can emerge even in small-world networks, with an infinite Hausdorff dimension, provided that the spectral dimension is finite. These results  reveal deep connections  between geometry and synchronization dynamics and are potentially very useful to further investigate the relation between structural and functional brain networks.

\section*{Acknowledgements}
We acknowledge interesting discussions with Z. Burda, R. Burioni, R. Loll, D. Mulder, L. Smolin, R. Sorkin, G. Vidal, and P. Jizba. 
We are  grateful for financial support from the Spanish Ministry of Science and the ``Agencia Espa\~nola de Investigaci\'on'' (AEI) under grant FIS2017-84256-P (FEDER funds) and from ``Obra Social La Caixa'' (ID 100010434, with code LCF/BQ/ES15/10360004). 
G.B. was partially supported by the  Perimeter Institute for Theoretical Physics (PI). The PI is supported by the Government of Canada through Industry Canada and by the Province of Ontario through the Ministry of Research and Innovation.

\appendix

\section{Stability of the  synchronized phase}
\label{a1}
In this Appendix we will investigate the stability of the synchronized phase by considering the linearized dynamical system given by Eqs. ($\ref{linearized}$).
The normalized Laplacian ${\bf {L}}$ appearing in Eqs. ($\ref{linearized}$) and defined in Eq. ($\ref{Laplacian}$) is diagonalizable with eigenvalues $\{\lambda_i\}_{i=1,2,\ldots, N}$, numbered in increasing order,  $0=\lambda_1<\lambda_2\leq \lambda_3,\ldots, \leq \lambda_N$, and therefore can be written as
\bea
\bf { P}^{-1} {L}P=D,
\eea
where ${\bf P}$ is the matrix whose columns are the right eigenvectors ${\bf v}^{\lambda}$ and
and ${\bf P}^{-1}$ is the matrix whose rows are the left eigenvectors ${\bf u}^{\lambda}$ of ${\bf {L}}$.
Notice that we always have ${\bf P}^{-1}{\bf P}={\bf I}$, where ${\bf I}$ indicates the identity matrix, due to the normalization condition of the eigenvectors given by Eq. ($\ref{norm}$).

The vector $\bm{\theta}=\left(\theta_1,\theta_2,\ldots, \theta_N\right)^{T}$ can be projected in the base of the right and left eigenvectors, so $\theta_i$ can be equivalently expressed as 
\bea
\theta_i&=& \sum_{\lambda}\theta_{\lambda}^R v_i^{\lambda},\nonumber \\
\theta_i&=&\sum_{\lambda} \theta_{\lambda}^L u_i^{\lambda},
\label{theta}
\eea
or, equivalently,
\bea
\bm{\theta}&=&{\bf P}\bm{\theta}^R,\nonumber \\
\bm{\theta}&=&[{\bf P}^{-1}]^{T}{\bm{\theta}}^L,
\label{theta_eigenvectors}
\eea
where  we have indicated with $\bm{\theta}^R$ and $ \bm{\theta}^L$ the column vector of elements $\theta_{\lambda}^R$ and $\theta_{\lambda}^L$, respectively. 
Inverting these relations we have that $\bm{\theta}^R$ and $\bm{\theta^L}$ are given by  
\bea
{\bm \theta}^{R}&=&{\bf P}^{-1} {\bm \theta},\nonumber \\
{\bm \theta}^{L}&=&{\bf P}^{T} {\bm \theta}.
\label{tRL}
\eea
Similarly we can also consider the vector $\bm{\omega}$ of elements $\omega_i$ and project it along the bases of the right and the left eigenvectors, 
\bea
\bm{\omega}&=&{\bf P}\bm{\omega}^R,\nonumber \\
\bm{\omega}&=&[{\bf P}^{-1}]^{T}{\bm{\omega}}^L.
\eea
Inverting these relations we obtain
\bea
{\bm \omega}^{R}&=&{\bf P}^{-1}{\bm \omega},\nonumber \\
{\bm \omega}^{L}&=&{\bf P}^{T}{\bm \omega}.
\eea
The linearized Eq. $(\ref{linearized})$ can also be projected  along the  bases of right and left eigenvectors getting 
\bea
\frac{d\theta^R_{\lambda}}{dt}&=&\omega^R_{\lambda}-\sigma \lambda \theta^R_{\lambda},\nonumber \\
\frac{d\theta^L_{\lambda}}{dt}&=&\omega^L_{\lambda}-\sigma \lambda \theta^L_{\lambda}.
\eea
This equations can be solved obtaining, for $\lambda\neq 0$,
\bea
\theta_{\lambda}^{R/L}(t)=e^{-\sigma\lambda t}\theta_{\lambda}^{R/L}(0)+\frac{\omega_{\lambda}^{R/L}}{\sigma \lambda}(1-e^{-\sigma\lambda t}),
\label{soln}
\eea
and, for $\lambda=0$,
\bea
\theta_{\lambda}^{R/L}(t)=\theta_{\lambda=0}^{R/L}(0)+{\omega_{\lambda=0}^{R/L}}t.
\label{sol0}
\eea

Finally, let  us  note that  $\bm{\omega}^{R,L}$ have the following averages 
\bea
\Avg{\omega^{R,L}_{\lambda}}&=&0.\nonumber \\
\Avg{\omega^{R}_{\lambda}\omega^{L}_{\lambda'}}&=&\sum_{i=1}^N \sum_{j=1}^N\Avg{\omega_i \omega_j} u_{i}^{\lambda}v_{j}^{\lambda'}=\delta_{\lambda,\lambda'}.
\label{omega_av}
\eea

As mentioned in the main text, in order  to evaluate the stability of the  synchronized state, we use an approach already established for finite lattices \cite{Choi,Chate} and  we calculate the average fluctuation of the phases over the entire network. These fluctuations are quantified by $W^2$ given by 
\bea
W^2=\frac{1}{N}\Avg{\sum_{i=1}^N[\theta_i(t)-\overline{\theta}]^2},
\eea
where  
\bea
\overline{\theta}= \frac{1}{N}\sum_{i=1}^N{\theta_i(t)}.
\eea
The divergence of  $W^2$  with the network size $N$ will indicate that the  synchronized phase is unstable.

Since $\bm{\theta}$ can be expressed equivalently in the base of right and left eigenvectors as expressed in Eqs. $(\ref{theta_eigenvectors})$, and the right eigenvector is given by the first of Eqs. $(\ref{eigen})$, we can calculate $\overline{\theta}$  in terms of $\bm{\theta}^L$ and $\bm{\theta}^R$ as 
\bea
\overline{\theta}&=&\sqrt{\frac{\avg{k}}{N}} {\theta}_{\lambda=0}^L(t)\nonumber \\
\overline{\theta}&=&\sum_{\lambda} \theta_{\lambda}^R(t) \frac{1}{N} \sum_i { v}_i^{\lambda}
\eea 
Using the explicit solution of $\theta^L_{\lambda}(t)$ and $\theta^R_{\lambda}(t)$ given by Eq. ($\ref{soln}$) and Eq. $(\ref{sol0})$  and using Eqs. (\ref{omega_av}) we can express  $\Avg{\overline{\theta}^2}$ as
\bea
\hspace*{-6mm}\Avg{\overline{\theta}^2}&=&\frac{1}{N}\Avg{\theta_{\lambda=0}^L (t)\theta_{\lambda=0}^R(t)}+ \sqrt{\frac{\avg{k}}{N}}\theta_{\lambda=0}^L (0)\theta_{\lambda=0}^R(0)\nonumber \\
&&\times\sum_{\lambda\neq 0}e^{-\sigma\lambda t}\frac{1}{N}\sum_{i=1}^Nv_i^{\lambda}.
\eea
Therefore asymptotically in time, for  $t\to \infty$, we obtain
\bea
\Avg{\overline{\theta}^2}=\frac{1}{N}\Avg{\theta_{\lambda=0}^L(t) \theta_{\lambda=0}^R(t)}.
\label{uno}
\eea
The fluctuations of the phases of the Kuramoto dynamics can be evaluated by considering that $W^2$ can be equivalently expressed as
\bea
W^2=\frac{1}{N}\Avg{\bm{\theta}^T\bm{\theta}}-\Avg{\overline{\theta}^2}.
\label{zero}
\eea
Using Eq. ($\ref{tRL}$) we note  that $\Avg{\bm{\theta}^T\bm{\theta}}$ has a simple expression in terms of $\bm{\theta}^L$ and $\bm{\theta}^R$, i.e.
\bea
\Avg{\bm{\theta}^T\bm{\theta}}=\Avg{[\bm{\theta}^{L}]^T{\bf P^{-1}}{\bf P}[\bm{\theta}^R]}=\Avg{[\bm{\theta}^L]^T\bm{\theta}^R}.
\eea
Using the solution of the Kuramoto dynamics Eq. $(\ref{soln})$ and Eqs. ($\ref{omega_av}$) we get
\bea
&&\Avg{[\bm{\theta}^L]^T\bm{\theta}^R}=\Avg{\theta_{\lambda=0}^L(t) \theta_{\lambda=0}^{R}(t)}\nonumber \\
&&\hspace*{-8mm}+\sum_{\{\lambda\}|\lambda\neq 0}\left[e^{-2\sigma\lambda t}\theta_{\lambda}^{R}(0)\theta_{\lambda}^{L}(0)+\frac{1}{(\sigma \lambda)^2}(1-e^{-\sigma\lambda t})^2\right].
\label{due}
\eea
Finally using Eq. $(\ref{zero})$ together with Eqs. ($\ref{uno}$)-($\ref{due}$), it results that asymptotically in time for  $t\to \infty$
\bea
W^2=\int_{\lambda_{2}}^{\lambda_{max}} d\lambda \rho(\lambda)\frac{1}{(\sigma \lambda)^2}.
\eea
Since the Fidler eigenvalue $\lambda_2$ satisfies the scaling expressed  in  Eq. $(\ref{Fidler})$ and goes to zero in the infinite network limit, using the scaling in Eq. ($\ref{scaling}$) for the density of eigenvalues $\rho(\lambda)$ we obtain the following results.
\begin{itemize}
\item[(1)]
For  spectral dimension $d_S<4$ the average fluctuation of the phases $W^2$ diverges as 
\bea
W^2\simeq O\left( \lambda_{2}^{d_S/2-2}\right)
\eea
\item[(2)] For spectral dimension $d_S=4$ the average fluctuation of the phases $W^2$ diverges as 
\bea
W^2 \simeq O( -\ln \lambda_{2}).
\eea
\item[(3)] Only for spectral dimension $d>4$ the average fluctuation of the phases
 $W^2$ converges.
 \end{itemize}
Specifically, by inserting the scaling of the Fidler eigenvalue Eq. $(\ref{Fidler})$ with the network size $N$  we obtain
\bea
     W^2 \sim  \left\{\begin{array}{ll}N ^{4/{d}_S-1} & \mbox{if }  {d}_S < 4 \\
		\ln(N) & \mbox{if } {d}_S = 4 \\
		\mbox{const} & \mbox{if } {d}_S > 4.
	\end{array}\right.
\eea
It follows from this derivation that the  synchronized state cannot be thermodynamically stable in networks with spectral dimension $d_S\leq 4$.

\section{Correlations between phases and validity of the linear approximation}
\label{a2}
In this Appendix we will evaluate the scaling of the  {\it correlation}  $C$ defined as 
\bea
    C= \frac{1}{N}\Avg{\bm{\theta}^{T} {\bf L}\bm{\theta}}
\eea
in the linear approximation. The divergence of the correlation $C$ in the large network limit indicates that the  linear approximation fails to be valid.
The correlation can be expressed in the basis of eigenvalues of the normalized Laplacian getting the simple expression
\bea
C=\frac{1}{N}\sum_{\lambda}\Avg{\theta_{\lambda}^{L}\lambda \theta_{\lambda}^R}.
\eea
By using the explicit expression for $\theta_{\lambda}^{L/R}$ given by Eq. $(\ref{soln})$ it is easy to show that 
\bea
C&=&\frac{1}{N}\sum_{\{\lambda\}|\lambda\neq 0}\lambda\left[e^{-2\sigma\lambda t}\theta_{\lambda}^{R}(0)\theta_{\lambda}^{L}(0)+\frac{1}{(\sigma \lambda)^2}(1-e^{-\sigma\lambda t})^2\right]\nonumber
\eea
which gives in the asymptotic limit $t\to \infty$
\bea
C&=& \int_{\lambda_2}^{\lambda_N} \rho(\lambda)\frac{1}{\sigma^2 \lambda} d\lambda.
\eea
By inserting the scaling  of the Fidler eigenvalue with the network size $N$ given by Eq. $(\ref{Fidler})$ we obtain
\bea
C\sim  \left\{
	\begin{array}{ll}
		 N ^{2/{d}_S-1} & \mbox{if }  {d}_S < 2 \\
		\ln(N) & \mbox{if } {d}_S = 2 \\
		\mbox{const} & \mbox{if } {d}_S > 2.
	\end{array}\right.
\eea
Therefore, for spectral dimension $d_S\leq 2$ the correlations among the phases of nearest neighbour nodes diverge and the linear approximation fails.

\section{Entrained phases}
\label{a3}
In this Appendix we will  characterize the fluctuations observed in phase velocities across the nodes of the network quantified by the global parameter $V^2$ given by 
\bea
V^2=\frac{1}{N}\sum_{i=1}^N\Avg{\left[{\psi}_i-\bar{\psi}\right]^2}
\eea
where $\psi_i$ indicates the phase velocity of node $i$ 
\bea
\psi_i=\dot{\theta_i},
\eea
and $\bar{\psi}$ the average of the phase velocities over the network
\bea
\bar{\psi}=\frac{1}{N}\sum_{i=0}^N\psi_i. 
\eea
The phase velocities $\bm{\psi}=(\psi_1,\psi_2,\ldots, \psi_N)^T$ can be projected into the basis of right and left eigenvectors of the normalized Laplacian getting 
\bea
{\bm \psi}^{R}&=&{\bf P}^{-1} {\bm \psi},\nonumber \\
{\bm \psi}^{L}&=&{\bf P}^{T} {\bm \psi}.
\label{psiRL}
\eea
By using the solution of the linearized dynamics, Eqs. ($\ref{soln}$) and $(\ref{sol0})$, it is easy to show that with the linear approximation we have
\bea
\psi_{\lambda}^{R/L}(t)=\dot{\theta_{\lambda}}^{R/L}=-\sigma \lambda e^{-\sigma\lambda t}\theta_{\lambda}^{R/L}(0)+{\omega_{\lambda}^{R/L}}e^{-\sigma\lambda t},
\label{vsoln}
\eea
and for $\lambda=0$
\bea
\psi_{\lambda}^{R/L}(t)={\omega_{\lambda=0}^{R/L}}.
\label{vsol0}
\eea
Using the same procedure used previously for the derivation of $\bar{\theta}$, it is easy to show that  the average phase velocity $\bar{\psi}$ can be expressed equivalently as 
\bea
\bar{\psi}&=& \sqrt{\frac{\avg{k}}{N}}{\psi}_{\lambda=0}^L(t),\nonumber \\
\bar{\psi}&=&\sum_{\lambda} \psi_{\lambda}^R(t) \frac{1}{N} \sum_i { v}_i^{\lambda}.
\eea 
From these expressions, and using Eqs. ($\ref{omega_av}$), it follows that 
\bea
\Avg{\bar{\psi}^2}=\frac{1}{N}\Avg{{\psi}_{\lambda=0}^L(t){\psi}_{\lambda=0}^r(t)}.
\eea
Finally using again Eq. (\ref{omega_av})  we get that 
\bea
{V^2}&=&\frac{1}{N}\Avg{[{\bm{\psi}^{L}}]^{T}{\bm{\psi}}^{R}}-\Avg{\bar{\psi}^2}
\eea
scales in the asymptotic limit $t\to \infty$ as
\bea
 V^2&\sim &\int_{\lambda_2}^{\lambda_{max}} d\lambda \rho(\lambda)e^{-2\sigma \lambda t}\nonumber \\ &\sim & t^{-d_S/2}.
\eea
This result implies that asymptotically in time the fluctuations in the phase velocities vanish, i.e.
\bea
V^2\to 0
\eea
as $t\to \infty$. This result implies that the phases of the oscillators are totally entrained as long as the linear approximation is valid.

\end{document}